\documentclass[aps, 10pt, amsmath, twocolumn, superscriptaddress, floatfix]{revtex4-1}
\usepackage[plainpages=false,colorlinks=true,linkcolor=blue, citecolor=blue, urlcolor=blue]{hyperref}
\usepackage[english]{babel}
\usepackage{graphicx}
\usepackage[charter]{mathdesign}
\usepackage{bm}
\usepackage[svgnames]{xcolor}

\newcommand\eva{\mathbf{e}_1}
\newcommand\evb{\mathbf{e}_2}

\newcommand\tv{\mathbf{t}}
\newcommand\kv{\mathbf{k}}

\newcommand\Gv{\mathbf{G}}

\newcommand\Sigmav{\bm{\Sigma}}
\newcommand\epsilonv{\bm{\epsilon}}
\newcommand\thetav{\bm{\theta}}
\newcommand\Gammav{\bm{\Gamma}}
\newcommand{\dg}{\dagger}
\newcommand{\fdag}{{\phantom{\dagger}}}

\newcommand\SCCO{Sr$_{14-x}$Ca$_x$Cu$_{24}$O$_{41}$}
\begin{document}
\title{Loop currents in ladder cuprates: A dynamical mean field theory study}
\author{Xiancong Lu}
\affiliation{Department of Physics, Xiamen University, Xiamen 361005, China}
\author{D. S\'en\'echal}
\affiliation{D\'epartement de physique and Institut quantique, Universit\'e de Sherbrooke, Sherbrooke, Qu\'ebec, Canada J1K 2R1}

\begin{abstract}
We investigate the possibility of spontaneous loop currents in the two-leg
ladder cuprate \SCCO\ by applying cluster dynamical mean field theory (CDMFT) to a seven-band Hubbard model for that compound, with 
an exact diagonalization solver.
We sample several values of the local interaction $U_d$ and of the Cu-O energy difference $E_{pd}$, by applying an external field
that induces loop currents. We find no instance of spontaneous loop currents once the external field is brought to zero.
\end{abstract}
\maketitle
\section{Introduction}

One of the most interesting features of cuprate superconductors is the pseudogap phenomenon, which is widely believed to be a key to understanding the mechanism of high-temperature superconductivity (HTSC)~\cite{ti.st.99}.
However, the origin of the pseudogap is still a matter of debate and the possibility of a spontaneously broken symmetry at low temperature within that state has not been excluded~\cite{ke.ki.15}.
One important possibility is the loop currents (LC) phase, proposed by Varma~\cite{varm.99,varm.06,sh.va.09}, in which equilibrium orbital currents are circulating along the O-Cu-O plaquette within each unit cell, thus breaking time-reversal symmetry while preserving translational symmetry.
Varma's proposal has stimulated many experimental searches for the signature of microscopic orbital magnetic moments.
Polarized neutron diffraction (PND) experiments have lent support to the existence of an intra-unit cell (IUC) magnetic order on CuO$_2$ planes~\cite{fa.si.06,mo.si.08,ma.li.17} or involving apical oxygens~\cite{li.ba.08}.
By contrast, nuclear magnetic resonance (NMR)~\cite{st.gr.11,mo.oh.13,wu.ma.15} and muon spin rotation ($\mu$SR)~\cite{ma.ac.08,so.pa.09,hu.pa.12,pa.ak.16} have not found evidence of magnetic order.
Varma's hypothesis has also been investigated theoretically, with numerical methods and models often used in the study of strongly correlated electrons, such as exact diagonalizations (ED)~\cite{gr.th.07, th.gr.08,ku.ch.14}, variational Monte Carlo (VMC)~\cite{we.la.09,we.gi.14}, and the variational cluster approximation (VCA)~\cite{lu.ch.12}.
For the three-band Hubbard model with realistic parameters for high-$T_c$ cuprates, the results of these different methods are consistent: the LC phase is not stabilized as a ground state in the thermodynamic limit.

The existence of LCs was also investigated theoretically in the two-leg ladder, which is simpler and interpolates between one- and two-dimensional systems.
By using the highly accurate density-matrix renormalization group (DMRG) technique, evidence for the existence of a
``staggered-flux'' phase was found for the two-leg ladder with long-range interaction, both at, and away from, half-filling~\cite{ma.fj.02,sc.ch.03}.
Analytical studies using a bosonization/renormalization group (RG) method also found stable regions of LCs for weak interactions~\cite{or.gi.97,ch.ga.08}.
However, a DMRG study on two-leg CuO ladders has found negative evidence towards the LC phase~\cite{ni.je.09,ni.je.10}.

Recently, using polarized neutron diffraction, Bounoua \textit{et al.}~\cite{bo.ma.09u} discovered the existence of a new kind of short-range magnetism in the two-leg ladder cuprate \SCCO (SCCO-$x$) for two Ca contents ($x=5$ and $x=8$).
The measured magnetic structure factor can be reproduced by assuming a set of counter-propagating LCs around each Cu atom.
This raises the possibility of a LC phase in the ladder cuprate.
In this paper, we try to verify this for $x=8$ using cluster dynamical mean field theory (CDMFT) applied on a multi-band Hubbard model.

\begin{figure*}[tp]
	\includegraphics[scale=0.55]{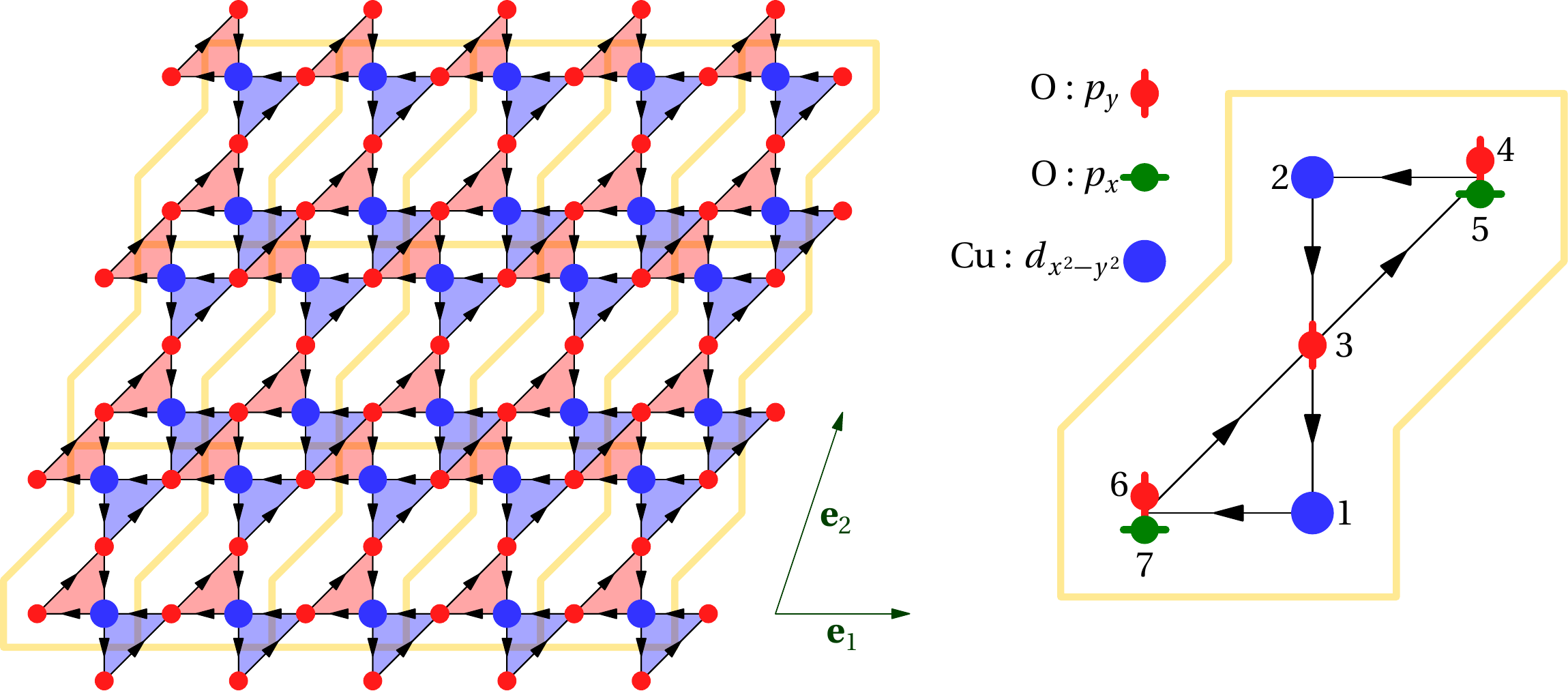}
	\caption{(Color online). Left panel: the Cu$_2$O$_3$ lattice. Cu atoms are in blue, oxygen atoms in red. The expected loop currents for SCCO-8 are shown by arrows and the associated fluxes of opposite signs are indicated by blue and red triangles.
	Unit cells are delimited in yellow and the lattice vectors $\eva$ and $\evb$ are shown.	Right panel: the seven orbitals in a given unit cell, with their labels as they appear in Eq.~\ref{eq:H0}.
	\label{fig:lattice}}
	\end{figure*}
	
\section{Model and method}\label{sec:intro}
\subsection{Hamiltonian}

The structure of SCCO-$x$ consists of an alternating stack of 1D CuO$_2$ chains and quasi-1D Cu$_2$O$_3$ two-leg ladder layers. 
We will focus of the ladder layer only and use a simplified description in terms of seven orbitals per unit cell: Two Cu $d_{x^2-y^2}$ orbitals (in blue on Fig.~\ref{fig:lattice}), two O $p_x$ and three O $p_y$ orbitals, respectively in green and red on the right panel of Fig.~\ref{fig:lattice}.
The hopping amplitudes will be chosen to be the same as the ones often used in the three-band model for the cuprates, except that two of the oxygen sites in the unit cell involve both $p_x$ and $p_y$ orbitals, owing to the slightly different geometry of the model compared to the cuprates. The noninteracting Hamiltonian has the form
\begin{equation}
H_0 = \sum_{\kv,\sigma} \tv_\kv  C_{\kv,\sigma}^\dg C_{\kv,\sigma}
\end{equation}
where $C_{\kv,\sigma}$ stands for an array of annihilation operators associated with the seven orbitals per unit cell, as labeled on Fig.~\ref{fig:lattice},
and where the momentum-dependent matrix $\tv_\kv$ is shown in Eq.~\eqref{eq:H0} below.
That matrix is Hermitian (the upper triangle is not shown).
The hopping amplitude between Cu and O orbitals is $t_{pd}$ and the energy difference between O and Cu orbitals is $E_{pd}$.
We assume for simplicity that the diagonal hopping amplitude $t_{pp}$ between oxygens is the same for $p_x-p_x$ and $p_x-p_y$ bonds.
We will set $t_{pp}=1$ and $t_{pd}=1.5$ throughout ($t_{pp}$ sets the energy scale).
Finally, we will specifically investigate the loop current structure illustrated on Fig.~\ref{fig:lattice} which, according to~\cite{bo.ma.09u}, is appropriate for $x=8$, corresponding to a doping of $\sim 17\%$.
We will, however, cover a fairly wide doping range around that value.
\begin{widetext}
\begin{equation}\label{eq:H0}
\begin{small}
\tv_\kv = 
-\begin{pmatrix}
0 &&&&&& \\
0 & 0 &&&&& \\
-t_{pd} & t_{pd} & -E_{pd} &&&& \\
t_{pd}e^{-i\kv\cdot\evb} & 0 & t_{pp}\left(1+e^{-i\kv\cdot\eva}\right) & -E_{pd} &&& \\
0 & t_{pd}\left(-1+e^{-i\kv\cdot\eva}\right) & t_{pp}\left(1-e^{-i\kv\cdot\eva}\right) & 0 & -E_{pd} && \\
0 & -t_{pd}e^{i\kv\cdot\evb} & 0 &  t_{pp}\left(1+e^{i\kv\cdot\eva}\right)e^{i\kv\cdot\evb} & t_{pp}\left(1-e^{i\kv\cdot\eva}\right)e^{i\kv\cdot\evb} & -E_{pd} & \\
t_{pd}\left(1-e^{i\kv\cdot\eva}\right) & 0 & t_{pp}\left(1-e^{i\kv\cdot\eva}\right) & t_{pp}\left(1-e^{i\kv\cdot\eva}\right)e^{i\kv\cdot\evb} & 0 & t_{pp}\left(1+e^{i\kv\cdot\eva}\right)e^{i\kv\cdot\evb} & -E_{pd} 
\end{pmatrix}
\end{small}
\end{equation}
\end{widetext}

To this noninteracting Hamiltonian we add local Hubbard interactions on the Cu and O atoms, so that the complete Hamiltonian reads
\begin{equation}\label{eq:H}
H = H_0 + U_d \sum_{i\in\mathrm{Cu}} n_{i\uparrow}^d n_{i\downarrow}^d + U_p \sum_{j\in\mathrm{O}} n_{j\uparrow}^p n_{j\downarrow}^p
-\mu \hat N_{\rm tot}
\end{equation}
where the sum over $i$ runs over Cu sites, the sum over $j$ runs over the five O orbitals in each unit cell, $\mu$ is the chemical potential and $\hat N$ the total number of electrons in all the orbitals considered.
$U_d$ and $U_p$ are the Coulomb repulsion of two holes sitting on the same copper orbital or the same oxygen orbital, respectively. 
We neglect the Coulomb interaction between different orbitals.

\begin{figure}[tp]
\includegraphics[scale=1.25]{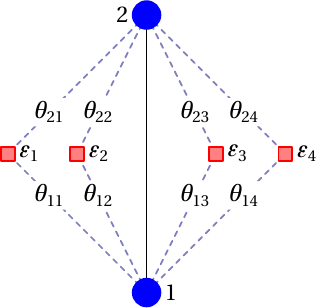}
\caption{(Color online) Structure of the hybridization between the four bath orbitals of the Anderson impurity model and the two Cu orbitals of the cluster (for simplicity, the oxygen orbitals are not shown, even though they are part of the impurity model).
\label{fig:bath}}
\end{figure}
		
\subsection{Impurity model}

In order to reveal loop currents possibly arising in model~\eqref{eq:H}, we use cluster dynamical mean-field theory (CDMFT)~\cite{li.ka.00,ko.sa.01,li.is.08,sene.15} with an exact diagonalization solver at zero temperature (or ED-CDMFT).
In CDMFT, the infinite lattice is tiled into identical units, or clusters, each of which is then coupled to a bath of uncorrelated, auxiliary orbitals.
The parameters describing this bath (energy levels, hybridization, etc.)
are then found by imposing a self-consistency condition. 

\begin{figure*}[tp]
\includegraphics[width=1.95\columnwidth]{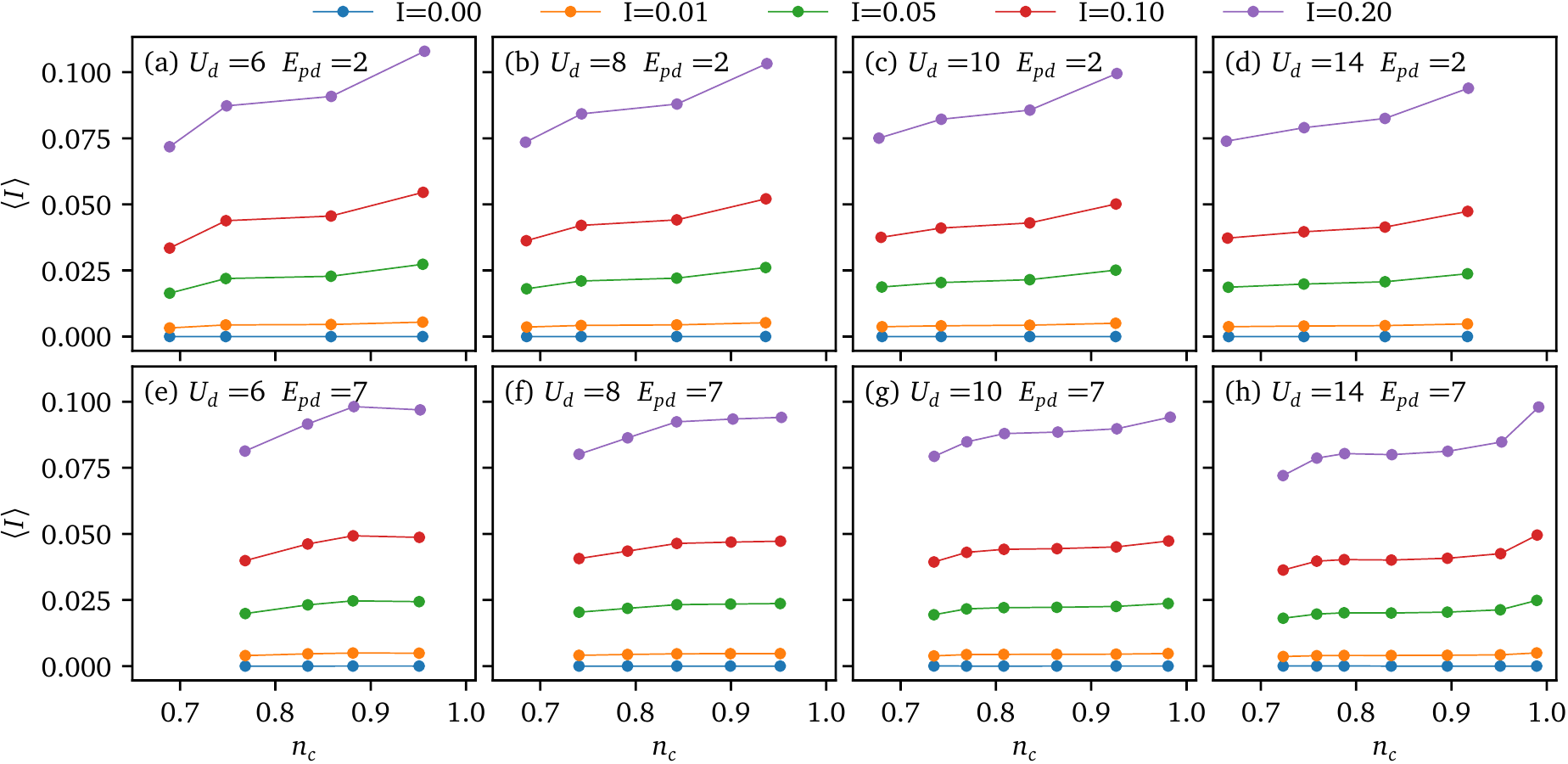}
\caption{(Color online). Expectation value $\langle \hat I\rangle$ of the local current operator as a function of electron density for several values of the external field $I$, for several values of $U_d$ and $E_{pd}$. In all cases $t_{pd}=1.5$, $t_{pp}=1$ and $U_p=3$. In the absence of external field ($I=0$), the loop current $\langle \hat I\rangle$ always vanishes.
\label{fig:I}}
\end{figure*}
			
In this work the cluster consists of a single unit cell (as shown on the right panel of Fig.~\ref{fig:lattice}), which is coupled to a bath of four uncorrelated orbitals.
The Cu orbitals being the most correlated (because $U_d$ is considerably larger than $U_p$), we choose a simplified bath parametrization in which the bath orbitals are hybridized with the Cu orbitals only, even though $U_p\ne0$, as shown on Fig.~\ref{fig:bath}. The corresponding \textit{Anderson impurity model} (AIM) Hamiltonian is
\begin{equation}\label{eq:Himp}
H_{\rm imp} = H_c + \sum_{i,r} \theta_{ir} \left(c_i^\dg a_r^\fdag + \mbox{H.c.} \right)
+ \sum_r \epsilon_{r} a_r^\dg a_{r}^\fdag,
\end{equation}
where $H_c$ is the Hamiltonian~\eqref{eq:H}, but restricted to a single cluster; cluster orbitals are labeled by the index $i$ and uncorrelated (bath) orbitals by the index $r$.
$\theta_{ir}$ is a complex hybridization parameter between cluster orbital $i$ and bath orbital $r$, and $\epsilon_{r}$ is the energy level of bath orbital $r$.
All these parameters are assumed to be spin independent, as we are not looking for magnetic ordering.

In ED-CDMFT, the bath parameters $\theta_{ir}$ and $\epsilon_r$ are determined by an approximate self-consistent procedure, as proposed initially in~\cite{ca.kr.94}, that goes as follows: (i) initial values $\{\epsilon_r, \theta_{ir}\}$ are chosen on the first iteration. (ii) For each iteration, the AIM \eqref{eq:Himp} is solved, i.e., the cluster Green function $\Gv_c(\omega)$ is computed using the Lanczos method.
The latter can be expressed as
\begin{equation}
\Gv_c(\omega)^{-1} = \omega - \tv_c - \Gammav(\omega) - \Sigmav_c(\omega)
\end{equation}
where $\tv_c$ is the one-body matrix in the cluster part of the impurity Hamiltonian $H_{\rm imp}$, $\Sigmav_c(\omega)$ is the associated self-energy, and $\Gammav(\omega)$ is the bath hybridization matrix:
\begin{equation}
\Gamma_{ij}(\omega) = \sum_{r}\frac{\theta_{ir}\theta_{jr}^*}{\omega - \epsilon_r}
\end{equation}
(iii) The bath parameters are updated, by minimizing the distance function:
\begin{equation}
d(\epsilonv, \thetav) = \sum_{i\omega_n} W(i\omega_n) \left[ \Gv_c(i\omega_n)^{-1} - \bar\Gv(i\omega_n)^{-1} \right]
\end{equation}
where $\bar\Gv(\omega)$, the projected Green function, is defined as
\begin{equation}\label{eq:GF}
\bar\Gv(\omega) = \frac{1}{N}\sum_\kv \Gv(\kv,\omega) \quad,\quad
\Gv(\kv,\omega) = \frac1{\omega - \tv_\kv - \Sigmav_c(\omega)}~~.
\end{equation}
In the above $\tv_\kv$ is the one-body Hamiltonian \eqref{eq:H0} and $N$ is the (nearly infinite) number of sites.
The matrices in the above are $7\times7$, for each spin projection.
Essentially, $\bar\Gv(\omega)$ is the local Green function obtained by carrying the self-energy $\Sigmav_c(\omega)$ to the whole lattice.
Ideally, $\bar\Gv(\omega)$ should coincide with the impurity Green function $\Gv_c(\omega)$, but the finite number of bath parameters does not allow for this correspondence at all frequencies, and so a distance function
$d(\epsilon_r, \theta_{ir})$ is defined, with emphasis on low frequencies along the imaginary axis.
The weight function $W(i\omega_n)$ is where the method has some arbitrariness; in this work $W(i\omega_n)$ is taken to be a constant for all Matsubara frequencies lower than a cutoff $\omega_c=2t_{pp}$, with a fictitious temperature
$\beta^{-1} = t_{pp}/50$. 
(iv) We go back to step (ii) and iterate until the bath parameters or the bath hybridization function
$\Gammav(\omega)$ stop varying within some preset tolerance.

\section{Results and discussion}\label{sec:res}

In order to probe the possible existence of loop currents in Model~\eqref{eq:H}, we first need to define an operator representing them.
We selected the following current loop operator, defined within the unit cell and following the arrows shown on the right panel of Fig.~\ref{fig:lattice}:
\begin{equation}
\hat I = i\Big(
c_1^\dg c_7 + c_7^\dg c_3 + c_3^\dg c_5 + c_5^\dg c_2 + c_2^\dg c_3  + c_3^\dg c_1
\Big)	+ \mathrm{H.c}
\end{equation}
We then impose an external field $I$ proportional to this operator on the system, i.e., we replace Hamiltonian \eqref{eq:H} by 
$H + I\hat I$.
This external field induces a nonzero expectation value $\langle \hat I\rangle$ on the impurity model.
We then reduce this external field to zero through a sequence of values (see Fig.~\ref{fig:I}) and monitor the expectation value $\langle \hat I\rangle$. 
If spontaneous currents were possible, a nonzero value of $\langle \hat I\rangle$ would persist down to $I=0$, which would indicate a spontaneous breaking of time reversal symmetry (TRS).
This is impossible if the hybridization $\theta_{ir}$ is purely real.
One can always require the hybridization parameter $\theta_{1r}$ to be real, because of an arbitrariness in the phase of the bath annihilation operator $a_{r\sigma}$. This being done, the phase of the other hybridization $\theta_{2r}$ is determined by the CDMFT procedure.
The complex-valued character of $\theta_{2r}$ is a necessary (but not sufficient) condition for a broken TRS state.

We have carried out a series of CDMFT computations on Model~\eqref{eq:H} with band parameters $t_{pp}=1$ and $t_{pd}=1.5$, $U_p=3$, and several values of $E_{pd}$ (0, 2, 4, 7), $U_d$ (6, 8, 10, 14) and chemical potential. 
In all cases an external current field $I$ was applied sequentially ($I=$0.2, 0.1, 0.05, 0.01 and 0.0) in order to maximize the chances of finding a spontaneous current.
In all cases no such current was found: $\langle\hat I\rangle=0$, within numerical error ($10^{-6}$).
Plots of $\langle\hat I\rangle$ vs the electron density $n_c$ on the impurity (corresponding to a few values of the chemical potential $\mu$) are shown on Fig.~\ref{fig:I}. In each panel the different curves correspond to different values of the external current field $I$, down to $I=0$ for the null curve. A few sample values of $U_d$ and $E_{pd}$ were chosen for the figure.

On Fig.~\ref{fig:lattice}, another current loop may be defined, that straddles four different unit cells, meeting at its center (dotted line on the figure). An operator $\hat I'$ exists for this current loop, except that it is defined on the lattice model only, not on the impurity.
Nevertheless, it is possible to formally compute the average of such an operator, from the lattice Green function $\Gv(\kv,\omega)$ of Eq.~\eqref{eq:GF}. We have checked that this average too is identically zero in the limit of zero external field.

We have also checked that our conclusions are unchanged if we add a sizable second-neighbor O-O hopping term $t_{pp}'=-1$. This 
hopping was deemed important to detect loop currents in Ref.~\cite{we.gi.14}. In our work, such a coupling does not affect the impurity model, but affects the CDMFT solution through the self-consistency solution.

Even though we are bound to limit ourselves to a sampling of model parameters, we are strongly inclined to conclude that spontaneous orbital currents do not occur in the model we used to describe \SCCO.
If the results of Ref.~\cite{bo.ma.09u} are truly the signature of loop currents, the source of the discrepancy has to be found either in the model itself, or in the simple CDMFT treatment we have put in place.
We have chosen an impurity model that contains a pair of triangular loops within the cluster, so as not to rely only on the measurement of lattice-based operators (as opposed to impurity-based operators). Of course the bath system itself is limited in size, but this is necessary in order to keep the problem numerically manageable. 
Increasing the number of bath orbitals would in general lead to better accuracy, but would not, from experience, change the nature of the ground state.
Quantum Monte Carlo studies are impossible here, because of the sign problem, which becomes a phase problem for complex-valued Hamiltonians.
In short, we do not believe that incremental improvements in the DMFT treatment of this problem would lead to different conclusions.

\begin{acknowledgments}
We thank A.-M. Tremblay for discussions and comments.
Computing resources were provided by Compute Canada and Calcul Qu\'ebec.
X.L. is supported by the National Natural Science Foundation of China (Grant No. 11974293) and the Fundamental Research Funds for Central Universities (Grant No. 20720180015).
\end{acknowledgments}

\begin{thebibliography}{34}%
	\makeatletter
	\providecommand \@ifxundefined [1]{%
		\@ifx{#1\undefined}
	}%
	\providecommand \@ifnum [1]{%
		\ifnum #1\expandafter \@firstoftwo
		\else \expandafter \@secondoftwo
		\fi
	}%
	\providecommand \@ifx [1]{%
		\ifx #1\expandafter \@firstoftwo
		\else \expandafter \@secondoftwo
		\fi
	}%
	\providecommand \natexlab [1]{#1}%
	\providecommand \enquote  [1]{``#1''}%
	\providecommand \bibnamefont  [1]{#1}%
	\providecommand \bibfnamefont [1]{#1}%
	\providecommand \citenamefont [1]{#1}%
	\providecommand \href@noop [0]{\@secondoftwo}%
	\providecommand \href [0]{\begingroup \@sanitize@url \@href}%
	\providecommand \@href[1]{\@@startlink{#1}\@@href}%
	\providecommand \@@href[1]{\endgroup#1\@@endlink}%
	\providecommand \@sanitize@url [0]{\catcode `\\12\catcode `\$12\catcode
			`\&12\catcode `\#12\catcode `\^12\catcode `\_12\catcode `\%12\relax}%
	\providecommand \@@startlink[1]{}%
	\providecommand \@@endlink[0]{}%
	\providecommand \url  [0]{\begingroup\@sanitize@url \@url }%
	\providecommand \@url [1]{\endgroup\@href {#1}{\urlprefix }}%
	\providecommand \urlprefix  [0]{URL }%
	\providecommand \Eprint [0]{\href }%
	\providecommand \doibase [0]{http://dx.doi.org/}%
	\providecommand \selectlanguage [0]{\@gobble}%
	\providecommand \bibinfo  [0]{\@secondoftwo}%
	\providecommand \bibfield  [0]{\@secondoftwo}%
	\providecommand \translation [1]{[#1]}%
	\providecommand \BibitemOpen [0]{}%
	\providecommand \bibitemStop [0]{}%
	\providecommand \bibitemNoStop [0]{.\EOS\space}%
	\providecommand \EOS [0]{\spacefactor3000\relax}%
	\providecommand \BibitemShut  [1]{\csname bibitem#1\endcsname}%
	\let\auto@bib@innerbib\@empty
	\bibitem [{\citenamefont {Timusk}\ and\ \citenamefont
			{Statt}(1999)}]{ti.st.99}%
			\BibitemOpen
			\bibfield  {author} {\bibinfo {author} {\bibfnamefont {T.}~\bibnamefont
			{Timusk}}\ and\ \bibinfo {author} {\bibfnamefont {B.}~\bibnamefont {Statt}},\
			}\href {\doibase 10.1088/0034-4885/62/1/002} {\bibfield  {journal} {\bibinfo
			{journal} {Reports on Progress in Physics}\ }\textbf {\bibinfo {volume}
			{62}},\ \bibinfo {pages} {61} (\bibinfo {year} {1999})}\BibitemShut {NoStop}%
	\bibitem [{\citenamefont {Keimer}\ \emph {et~al.}(2015)\citenamefont {Keimer},
			\citenamefont {Kivelson}, \citenamefont {Norman}, \citenamefont {Uchida},\
			and\ \citenamefont {Zaanen}}]{ke.ki.15}%
			\BibitemOpen
			\bibfield  {author} {\bibinfo {author} {\bibfnamefont {B.}~\bibnamefont
			{Keimer}}, \bibinfo {author} {\bibfnamefont {S.~A.}\ \bibnamefont
			{Kivelson}}, \bibinfo {author} {\bibfnamefont {M.~R.}\ \bibnamefont
			{Norman}}, \bibinfo {author} {\bibfnamefont {S.}~\bibnamefont {Uchida}}, \
			and\ \bibinfo {author} {\bibfnamefont {J.}~\bibnamefont {Zaanen}},\ }\href
			{\doibase 10.1038/nature14165} {\bibfield  {journal} {\bibinfo  {journal}
			{Nature}\ }\textbf {\bibinfo {volume} {518}},\ \bibinfo {pages} {179}
			(\bibinfo {year} {2015})}\BibitemShut {NoStop}%
	\bibitem [{\citenamefont {Varma}(1999)}]{varm.99}%
			\BibitemOpen
			\bibfield  {author} {\bibinfo {author} {\bibfnamefont {C.~M.}\ \bibnamefont
			{Varma}},\ }\href {\doibase 10.1103/PhysRevLett.83.3538} {\bibfield
			{journal} {\bibinfo  {journal} {Phys. Rev. Lett.}\ }\textbf {\bibinfo
			{volume} {83}},\ \bibinfo {pages} {3538} (\bibinfo {year}
			{1999})}\BibitemShut {NoStop}%
	\bibitem [{\citenamefont {Varma}(2006)}]{varm.06}%
			\BibitemOpen
			\bibfield  {author} {\bibinfo {author} {\bibfnamefont {C.~M.}\ \bibnamefont
			{Varma}},\ }\href {\doibase 10.1103/PhysRevB.73.155113} {\bibfield  {journal}
			{\bibinfo  {journal} {Phys. Rev. B}\ }\textbf {\bibinfo {volume} {73}},\
			\bibinfo {pages} {155113} (\bibinfo {year} {2006})}\BibitemShut {NoStop}%
	\bibitem [{\citenamefont {Shekhter}\ and\ \citenamefont
			{Varma}(2009)}]{sh.va.09}%
			\BibitemOpen
			\bibfield  {author} {\bibinfo {author} {\bibfnamefont {A.}~\bibnamefont
			{Shekhter}}\ and\ \bibinfo {author} {\bibfnamefont {C.~M.}\ \bibnamefont
			{Varma}},\ }\href {\doibase 10.1103/PhysRevB.80.214501} {\bibfield  {journal}
			{\bibinfo  {journal} {Phys. Rev. B}\ }\textbf {\bibinfo {volume} {80}},\
			\bibinfo {pages} {214501} (\bibinfo {year} {2009})}\BibitemShut {NoStop}%
	\bibitem [{\citenamefont {Fauqu\'e}\ \emph {et~al.}(2006)\citenamefont
			{Fauqu\'e}, \citenamefont {Sidis}, \citenamefont {Hinkov}, \citenamefont
			{Pailh\`es}, \citenamefont {Lin}, \citenamefont {Chaud},\ and\ \citenamefont
			{Bourges}}]{fa.si.06}%
			\BibitemOpen
			\bibfield  {author} {\bibinfo {author} {\bibfnamefont {B.}~\bibnamefont
			{Fauqu\'e}}, \bibinfo {author} {\bibfnamefont {Y.}~\bibnamefont {Sidis}},
			\bibinfo {author} {\bibfnamefont {V.}~\bibnamefont {Hinkov}}, \bibinfo
			{author} {\bibfnamefont {S.}~\bibnamefont {Pailh\`es}}, \bibinfo {author}
			{\bibfnamefont {C.~T.}\ \bibnamefont {Lin}}, \bibinfo {author} {\bibfnamefont
			{X.}~\bibnamefont {Chaud}}, \ and\ \bibinfo {author} {\bibfnamefont
			{P.}~\bibnamefont {Bourges}},\ }\href {\doibase
			10.1103/PhysRevLett.96.197001} {\bibfield  {journal} {\bibinfo  {journal}
			{Phys. Rev. Lett.}\ }\textbf {\bibinfo {volume} {96}},\ \bibinfo {pages}
			{197001} (\bibinfo {year} {2006})}\BibitemShut {NoStop}%
	\bibitem [{\citenamefont {Mook}\ \emph {et~al.}(2008)\citenamefont {Mook},
			\citenamefont {Sidis}, \citenamefont {Fauqu\'e}, \citenamefont {Bal\'edent},\
			and\ \citenamefont {Bourges}}]{mo.si.08}%
			\BibitemOpen
			\bibfield  {author} {\bibinfo {author} {\bibfnamefont {H.~A.}\ \bibnamefont
			{Mook}}, \bibinfo {author} {\bibfnamefont {Y.}~\bibnamefont {Sidis}},
			\bibinfo {author} {\bibfnamefont {B.}~\bibnamefont {Fauqu\'e}}, \bibinfo
			{author} {\bibfnamefont {V.}~\bibnamefont {Bal\'edent}}, \ and\ \bibinfo
			{author} {\bibfnamefont {P.}~\bibnamefont {Bourges}},\ }\href {\doibase
			10.1103/PhysRevB.78.020506} {\bibfield  {journal} {\bibinfo  {journal} {Phys.
			Rev. B}\ }\textbf {\bibinfo {volume} {78}},\ \bibinfo {pages} {020506}
			(\bibinfo {year} {2008})}\BibitemShut {NoStop}%
	\bibitem [{\citenamefont {Mangin-Thro}\ \emph {et~al.}(2017)\citenamefont
			{Mangin-Thro}, \citenamefont {Li}, \citenamefont {Sidis},\ and\ \citenamefont
			{Bourges}}]{ma.li.17}%
			\BibitemOpen
			\bibfield  {author} {\bibinfo {author} {\bibfnamefont {L.}~\bibnamefont
			{Mangin-Thro}}, \bibinfo {author} {\bibfnamefont {Y.}~\bibnamefont {Li}},
			\bibinfo {author} {\bibfnamefont {Y.}~\bibnamefont {Sidis}}, \ and\ \bibinfo
			{author} {\bibfnamefont {P.}~\bibnamefont {Bourges}},\ }\href {\doibase
			10.1103/PhysRevLett.118.097003} {\bibfield  {journal} {\bibinfo  {journal}
			{Phys. Rev. Lett.}\ }\textbf {\bibinfo {volume} {118}},\ \bibinfo {pages}
			{097003} (\bibinfo {year} {2017})}\BibitemShut {NoStop}%
	\bibitem [{\citenamefont {Li}\ \emph {et~al.}(2008)\citenamefont {Li},
			\citenamefont {Bal{\'e}dent}, \citenamefont {Bari{\v s}i{\'c}}, \citenamefont
			{Cho}, \citenamefont {Fauqu{\'e}}, \citenamefont {Sidis}, \citenamefont {Yu},
			\citenamefont {Zhao}, \citenamefont {Bourges},\ and\ \citenamefont
			{Greven}}]{li.ba.08}%
			\BibitemOpen
			\bibfield  {author} {\bibinfo {author} {\bibfnamefont {Y.}~\bibnamefont
			{Li}}, \bibinfo {author} {\bibfnamefont {V.}~\bibnamefont {Bal{\'e}dent}},
			\bibinfo {author} {\bibfnamefont {N.}~\bibnamefont {Bari{\v s}i{\'c}}},
			\bibinfo {author} {\bibfnamefont {Y.}~\bibnamefont {Cho}}, \bibinfo {author}
			{\bibfnamefont {B.}~\bibnamefont {Fauqu{\'e}}}, \bibinfo {author}
			{\bibfnamefont {Y.}~\bibnamefont {Sidis}}, \bibinfo {author} {\bibfnamefont
			{G.}~\bibnamefont {Yu}}, \bibinfo {author} {\bibfnamefont {X.}~\bibnamefont
			{Zhao}}, \bibinfo {author} {\bibfnamefont {P.}~\bibnamefont {Bourges}}, \
			and\ \bibinfo {author} {\bibfnamefont {M.}~\bibnamefont {Greven}},\ }\href
			{\doibase 10.1038/nature07251} {\bibfield  {journal} {\bibinfo  {journal}
			{Nature}\ }\textbf {\bibinfo {volume} {455}},\ \bibinfo {pages} {372}
			(\bibinfo {year} {2008})}\BibitemShut {NoStop}%
	\bibitem [{\citenamefont {Str\"assle}\ \emph {et~al.}(2011)\citenamefont
			{Str\"assle}, \citenamefont {Graneli}, \citenamefont {Mali}, \citenamefont
			{Roos},\ and\ \citenamefont {Keller}}]{st.gr.11}%
			\BibitemOpen
			\bibfield  {author} {\bibinfo {author} {\bibfnamefont {S.}~\bibnamefont
			{Str\"assle}}, \bibinfo {author} {\bibfnamefont {B.}~\bibnamefont {Graneli}},
			\bibinfo {author} {\bibfnamefont {M.}~\bibnamefont {Mali}}, \bibinfo {author}
			{\bibfnamefont {J.}~\bibnamefont {Roos}}, \ and\ \bibinfo {author}
			{\bibfnamefont {H.}~\bibnamefont {Keller}},\ }\href {\doibase
			10.1103/PhysRevLett.106.097003} {\bibfield  {journal} {\bibinfo  {journal}
			{Phys. Rev. Lett.}\ }\textbf {\bibinfo {volume} {106}},\ \bibinfo {pages}
			{097003} (\bibinfo {year} {2011})}\BibitemShut {NoStop}%
	\bibitem [{\citenamefont {Mounce}\ \emph {et~al.}(2013)\citenamefont {Mounce},
			\citenamefont {Oh}, \citenamefont {Lee}, \citenamefont {Halperin},
			\citenamefont {Reyes}, \citenamefont {Kuhns}, \citenamefont {Chan},
			\citenamefont {Dorow}, \citenamefont {Ji}, \citenamefont {Xia}, \citenamefont
			{Zhao},\ and\ \citenamefont {Greven}}]{mo.oh.13}%
			\BibitemOpen
			\bibfield  {author} {\bibinfo {author} {\bibfnamefont {A.~M.}\ \bibnamefont
			{Mounce}}, \bibinfo {author} {\bibfnamefont {S.}~\bibnamefont {Oh}}, \bibinfo
			{author} {\bibfnamefont {J.~A.}\ \bibnamefont {Lee}}, \bibinfo {author}
			{\bibfnamefont {W.~P.}\ \bibnamefont {Halperin}}, \bibinfo {author}
			{\bibfnamefont {A.~P.}\ \bibnamefont {Reyes}}, \bibinfo {author}
			{\bibfnamefont {P.~L.}\ \bibnamefont {Kuhns}}, \bibinfo {author}
			{\bibfnamefont {M.~K.}\ \bibnamefont {Chan}}, \bibinfo {author}
			{\bibfnamefont {C.}~\bibnamefont {Dorow}}, \bibinfo {author} {\bibfnamefont
			{L.}~\bibnamefont {Ji}}, \bibinfo {author} {\bibfnamefont {D.}~\bibnamefont
			{Xia}}, \bibinfo {author} {\bibfnamefont {X.}~\bibnamefont {Zhao}}, \ and\
			\bibinfo {author} {\bibfnamefont {M.}~\bibnamefont {Greven}},\ }\href
			{\doibase 10.1103/PhysRevLett.111.187003} {\bibfield  {journal} {\bibinfo
			{journal} {Phys. Rev. Lett.}\ }\textbf {\bibinfo {volume} {111}},\ \bibinfo
			{pages} {187003} (\bibinfo {year} {2013})}\BibitemShut {NoStop}%
	\bibitem [{\citenamefont {Wu}\ \emph {et~al.}(2015)\citenamefont {Wu},
			\citenamefont {Mayaffre}, \citenamefont {Kr{\"a}mer}, \citenamefont
			{Horvati{\'c}}, \citenamefont {Berthier}, \citenamefont {Hardy},
			\citenamefont {Liang}, \citenamefont {Bonn},\ and\ \citenamefont
			{Julien}}]{wu.ma.15}%
			\BibitemOpen
			\bibfield  {author} {\bibinfo {author} {\bibfnamefont {T.}~\bibnamefont
			{Wu}}, \bibinfo {author} {\bibfnamefont {H.}~\bibnamefont {Mayaffre}},
			\bibinfo {author} {\bibfnamefont {S.}~\bibnamefont {Kr{\"a}mer}}, \bibinfo
			{author} {\bibfnamefont {M.}~\bibnamefont {Horvati{\'c}}}, \bibinfo {author}
			{\bibfnamefont {C.}~\bibnamefont {Berthier}}, \bibinfo {author}
			{\bibfnamefont {W.~N.}\ \bibnamefont {Hardy}}, \bibinfo {author}
			{\bibfnamefont {R.}~\bibnamefont {Liang}}, \bibinfo {author} {\bibfnamefont
			{D.~A.}\ \bibnamefont {Bonn}}, \ and\ \bibinfo {author} {\bibfnamefont
			{M.-H.}\ \bibnamefont {Julien}},\ }\href {\doibase 10.1038/ncomms7438}
			{\bibfield  {journal} {\bibinfo  {journal} {Nature Communications}\ }\textbf
			{\bibinfo {volume} {6}},\ \bibinfo {pages} {6438} (\bibinfo {year}
			{2015})}\BibitemShut {NoStop}%
	\bibitem [{\citenamefont {MacDougall}\ \emph {et~al.}(2008)\citenamefont
			{MacDougall}, \citenamefont {Aczel}, \citenamefont {Carlo}, \citenamefont
			{Ito}, \citenamefont {Rodriguez}, \citenamefont {Russo}, \citenamefont
			{Uemura}, \citenamefont {Wakimoto},\ and\ \citenamefont {Luke}}]{ma.ac.08}%
			\BibitemOpen
			\bibfield  {author} {\bibinfo {author} {\bibfnamefont {G.~J.}\ \bibnamefont
			{MacDougall}}, \bibinfo {author} {\bibfnamefont {A.~A.}\ \bibnamefont
			{Aczel}}, \bibinfo {author} {\bibfnamefont {J.~P.}\ \bibnamefont {Carlo}},
			\bibinfo {author} {\bibfnamefont {T.}~\bibnamefont {Ito}}, \bibinfo {author}
			{\bibfnamefont {J.}~\bibnamefont {Rodriguez}}, \bibinfo {author}
			{\bibfnamefont {P.~L.}\ \bibnamefont {Russo}}, \bibinfo {author}
			{\bibfnamefont {Y.~J.}\ \bibnamefont {Uemura}}, \bibinfo {author}
			{\bibfnamefont {S.}~\bibnamefont {Wakimoto}}, \ and\ \bibinfo {author}
			{\bibfnamefont {G.~M.}\ \bibnamefont {Luke}},\ }\href {\doibase
			10.1103/PhysRevLett.101.017001} {\bibfield  {journal} {\bibinfo  {journal}
			{Phys. Rev. Lett.}\ }\textbf {\bibinfo {volume} {101}},\ \bibinfo {pages}
			{017001} (\bibinfo {year} {2008})}\BibitemShut {NoStop}%
	\bibitem [{\citenamefont {Sonier}\ \emph {et~al.}(2009)\citenamefont {Sonier},
			\citenamefont {Pacradouni}, \citenamefont {Sabok-Sayr}, \citenamefont
			{Hardy}, \citenamefont {Bonn}, \citenamefont {Liang},\ and\ \citenamefont
			{Mook}}]{so.pa.09}%
			\BibitemOpen
			\bibfield  {author} {\bibinfo {author} {\bibfnamefont {J.~E.}\ \bibnamefont
			{Sonier}}, \bibinfo {author} {\bibfnamefont {V.}~\bibnamefont {Pacradouni}},
			\bibinfo {author} {\bibfnamefont {S.~A.}\ \bibnamefont {Sabok-Sayr}},
			\bibinfo {author} {\bibfnamefont {W.~N.}\ \bibnamefont {Hardy}}, \bibinfo
			{author} {\bibfnamefont {D.~A.}\ \bibnamefont {Bonn}}, \bibinfo {author}
			{\bibfnamefont {R.}~\bibnamefont {Liang}}, \ and\ \bibinfo {author}
			{\bibfnamefont {H.~A.}\ \bibnamefont {Mook}},\ }\href {\doibase
			10.1103/PhysRevLett.103.167002} {\bibfield  {journal} {\bibinfo  {journal}
			{Phys. Rev. Lett.}\ }\textbf {\bibinfo {volume} {103}},\ \bibinfo {pages}
			{167002} (\bibinfo {year} {2009})}\BibitemShut {NoStop}%
	\bibitem [{\citenamefont {Huang}\ \emph {et~al.}(2012)\citenamefont {Huang},
			\citenamefont {Pacradouni}, \citenamefont {Kennett}, \citenamefont {Komiya},\
			and\ \citenamefont {Sonier}}]{hu.pa.12}%
			\BibitemOpen
			\bibfield  {author} {\bibinfo {author} {\bibfnamefont {W.}~\bibnamefont
			{Huang}}, \bibinfo {author} {\bibfnamefont {V.}~\bibnamefont {Pacradouni}},
			\bibinfo {author} {\bibfnamefont {M.~P.}\ \bibnamefont {Kennett}}, \bibinfo
			{author} {\bibfnamefont {S.}~\bibnamefont {Komiya}}, \ and\ \bibinfo {author}
			{\bibfnamefont {J.~E.}\ \bibnamefont {Sonier}},\ }\href {\doibase
			10.1103/PhysRevB.85.104527} {\bibfield  {journal} {\bibinfo  {journal} {Phys.
			Rev. B}\ }\textbf {\bibinfo {volume} {85}},\ \bibinfo {pages} {104527}
			(\bibinfo {year} {2012})}\BibitemShut {NoStop}%
	\bibitem [{\citenamefont {Pal}\ \emph {et~al.}(2016)\citenamefont {Pal},
			\citenamefont {Akintola}, \citenamefont {Potma}, \citenamefont {Ishikado},
			\citenamefont {Eisaki}, \citenamefont {Hardy}, \citenamefont {Bonn},
			\citenamefont {Liang},\ and\ \citenamefont {Sonier}}]{pa.ak.16}%
			\BibitemOpen
			\bibfield  {author} {\bibinfo {author} {\bibfnamefont {A.}~\bibnamefont
			{Pal}}, \bibinfo {author} {\bibfnamefont {K.}~\bibnamefont {Akintola}},
			\bibinfo {author} {\bibfnamefont {M.}~\bibnamefont {Potma}}, \bibinfo
			{author} {\bibfnamefont {M.}~\bibnamefont {Ishikado}}, \bibinfo {author}
			{\bibfnamefont {H.}~\bibnamefont {Eisaki}}, \bibinfo {author} {\bibfnamefont
			{W.~N.}\ \bibnamefont {Hardy}}, \bibinfo {author} {\bibfnamefont {D.~A.}\
			\bibnamefont {Bonn}}, \bibinfo {author} {\bibfnamefont {R.}~\bibnamefont
			{Liang}}, \ and\ \bibinfo {author} {\bibfnamefont {J.~E.}\ \bibnamefont
			{Sonier}},\ }\href {\doibase 10.1103/PhysRevB.94.134514} {\bibfield
			{journal} {\bibinfo  {journal} {Phys. Rev. B}\ }\textbf {\bibinfo {volume}
			{94}},\ \bibinfo {pages} {134514} (\bibinfo {year} {2016})}\BibitemShut
			{NoStop}%
	\bibitem [{\citenamefont {Greiter}\ and\ \citenamefont
			{Thomale}(2007)}]{gr.th.07}%
			\BibitemOpen
			\bibfield  {author} {\bibinfo {author} {\bibfnamefont {M.}~\bibnamefont
			{Greiter}}\ and\ \bibinfo {author} {\bibfnamefont {R.}~\bibnamefont
			{Thomale}},\ }\href {\doibase 10.1103/PhysRevLett.99.027005} {\bibfield
			{journal} {\bibinfo  {journal} {Phys. Rev. Lett.}\ }\textbf {\bibinfo
			{volume} {99}},\ \bibinfo {pages} {027005} (\bibinfo {year}
			{2007})}\BibitemShut {NoStop}%
	\bibitem [{\citenamefont {Thomale}\ and\ \citenamefont
			{Greiter}(2008)}]{th.gr.08}%
			\BibitemOpen
			\bibfield  {author} {\bibinfo {author} {\bibfnamefont {R.}~\bibnamefont
			{Thomale}}\ and\ \bibinfo {author} {\bibfnamefont {M.}~\bibnamefont
			{Greiter}},\ }\href {\doibase 10.1103/PhysRevB.77.094511} {\bibfield
			{journal} {\bibinfo  {journal} {Phys. Rev. B}\ }\textbf {\bibinfo {volume}
			{77}},\ \bibinfo {pages} {094511} (\bibinfo {year} {2008})}\BibitemShut
			{NoStop}%
	\bibitem [{\citenamefont {Kung}\ \emph {et~al.}(2014)\citenamefont {Kung},
			\citenamefont {Chen}, \citenamefont {Moritz}, \citenamefont {Johnston},
			\citenamefont {Thomale},\ and\ \citenamefont {Devereaux}}]{ku.ch.14}%
			\BibitemOpen
			\bibfield  {author} {\bibinfo {author} {\bibfnamefont {Y.~F.}\ \bibnamefont
			{Kung}}, \bibinfo {author} {\bibfnamefont {C.-C.}\ \bibnamefont {Chen}},
			\bibinfo {author} {\bibfnamefont {B.}~\bibnamefont {Moritz}}, \bibinfo
			{author} {\bibfnamefont {S.}~\bibnamefont {Johnston}}, \bibinfo {author}
			{\bibfnamefont {R.}~\bibnamefont {Thomale}}, \ and\ \bibinfo {author}
			{\bibfnamefont {T.~P.}\ \bibnamefont {Devereaux}},\ }\href {\doibase
			10.1103/PhysRevB.90.224507} {\bibfield  {journal} {\bibinfo  {journal} {Phys.
			Rev. B}\ }\textbf {\bibinfo {volume} {90}},\ \bibinfo {pages} {224507}
			(\bibinfo {year} {2014})}\BibitemShut {NoStop}%
	\bibitem [{\citenamefont {Weber}\ \emph {et~al.}(2009)\citenamefont {Weber},
			\citenamefont {L\"auchli}, \citenamefont {Mila},\ and\ \citenamefont
			{Giamarchi}}]{we.la.09}%
			\BibitemOpen
			\bibfield  {author} {\bibinfo {author} {\bibfnamefont {C.}~\bibnamefont
			{Weber}}, \bibinfo {author} {\bibfnamefont {A.}~\bibnamefont {L\"auchli}},
			\bibinfo {author} {\bibfnamefont {F.}~\bibnamefont {Mila}}, \ and\ \bibinfo
			{author} {\bibfnamefont {T.}~\bibnamefont {Giamarchi}},\ }\href {\doibase
			10.1103/PhysRevLett.102.017005} {\bibfield  {journal} {\bibinfo  {journal}
			{Phys. Rev. Lett.}\ }\textbf {\bibinfo {volume} {102}},\ \bibinfo {pages}
			{017005} (\bibinfo {year} {2009})}\BibitemShut {NoStop}%
	\bibitem [{\citenamefont {Weber}\ \emph {et~al.}(2014)\citenamefont {Weber},
			\citenamefont {Giamarchi},\ and\ \citenamefont {Varma}}]{we.gi.14}%
			\BibitemOpen
			\bibfield  {author} {\bibinfo {author} {\bibfnamefont {C.}~\bibnamefont
			{Weber}}, \bibinfo {author} {\bibfnamefont {T.}~\bibnamefont {Giamarchi}}, \
			and\ \bibinfo {author} {\bibfnamefont {C.~M.}\ \bibnamefont {Varma}},\ }\href
			{\doibase 10.1103/PhysRevLett.112.117001} {\bibfield  {journal} {\bibinfo
			{journal} {Phys. Rev. Lett.}\ }\textbf {\bibinfo {volume} {112}},\ \bibinfo
			{pages} {117001} (\bibinfo {year} {2014})}\BibitemShut {NoStop}%
	\bibitem [{\citenamefont {Lu}\ \emph {et~al.}(2012)\citenamefont {Lu},
			\citenamefont {Chioncel},\ and\ \citenamefont {Arrigoni}}]{lu.ch.12}%
			\BibitemOpen
			\bibfield  {author} {\bibinfo {author} {\bibfnamefont {X.}~\bibnamefont
			{Lu}}, \bibinfo {author} {\bibfnamefont {L.}~\bibnamefont {Chioncel}}, \ and\
			\bibinfo {author} {\bibfnamefont {E.}~\bibnamefont {Arrigoni}},\ }\href
			{\doibase 10.1103/PhysRevB.85.125117} {\bibfield  {journal} {\bibinfo
			{journal} {Phys. Rev. B}\ }\textbf {\bibinfo {volume} {85}},\ \bibinfo
			{pages} {125117} (\bibinfo {year} {2012})}\BibitemShut {NoStop}%
	\bibitem [{\citenamefont {Marston}\ \emph {et~al.}(2002)\citenamefont
			{Marston}, \citenamefont {Fj\ae{}restad},\ and\ \citenamefont
			{Sudb\o{}}}]{ma.fj.02}%
			\BibitemOpen
			\bibfield  {author} {\bibinfo {author} {\bibfnamefont {J.~B.}\ \bibnamefont
			{Marston}}, \bibinfo {author} {\bibfnamefont {J.~O.}\ \bibnamefont
			{Fj\ae{}restad}}, \ and\ \bibinfo {author} {\bibfnamefont {A.}~\bibnamefont
			{Sudb\o{}}},\ }\href {\doibase 10.1103/PhysRevLett.89.056404} {\bibfield
			{journal} {\bibinfo  {journal} {Phys. Rev. Lett.}\ }\textbf {\bibinfo
			{volume} {89}},\ \bibinfo {pages} {056404} (\bibinfo {year}
			{2002})}\BibitemShut {NoStop}%
	\bibitem [{\citenamefont {Schollw\"ock}\ \emph {et~al.}(2003)\citenamefont
			{Schollw\"ock}, \citenamefont {Chakravarty}, \citenamefont {Fj\ae{}restad},
			\citenamefont {Marston},\ and\ \citenamefont {Troyer}}]{sc.ch.03}%
			\BibitemOpen
			\bibfield  {author} {\bibinfo {author} {\bibfnamefont {U.}~\bibnamefont
			{Schollw\"ock}}, \bibinfo {author} {\bibfnamefont {S.}~\bibnamefont
			{Chakravarty}}, \bibinfo {author} {\bibfnamefont {J.~O.}\ \bibnamefont
			{Fj\ae{}restad}}, \bibinfo {author} {\bibfnamefont {J.~B.}\ \bibnamefont
			{Marston}}, \ and\ \bibinfo {author} {\bibfnamefont {M.}~\bibnamefont
			{Troyer}},\ }\href {\doibase 10.1103/PhysRevLett.90.186401} {\bibfield
			{journal} {\bibinfo  {journal} {Phys. Rev. Lett.}\ }\textbf {\bibinfo
			{volume} {90}},\ \bibinfo {pages} {186401} (\bibinfo {year}
			{2003})}\BibitemShut {NoStop}%
	\bibitem [{\citenamefont {Orignac}\ and\ \citenamefont
			{Giamarchi}(1997)}]{or.gi.97}%
			\BibitemOpen
			\bibfield  {author} {\bibinfo {author} {\bibfnamefont {E.}~\bibnamefont
			{Orignac}}\ and\ \bibinfo {author} {\bibfnamefont {T.}~\bibnamefont
			{Giamarchi}},\ }\href {\doibase 10.1103/PhysRevB.56.7167} {\bibfield
			{journal} {\bibinfo  {journal} {Phys. Rev. B}\ }\textbf {\bibinfo {volume}
			{56}},\ \bibinfo {pages} {7167} (\bibinfo {year} {1997})}\BibitemShut
			{NoStop}%
	\bibitem [{\citenamefont {Chudzinski}\ \emph {et~al.}(2008)\citenamefont
			{Chudzinski}, \citenamefont {Gabay},\ and\ \citenamefont
			{Giamarchi}}]{ch.ga.08}%
			\BibitemOpen
			\bibfield  {author} {\bibinfo {author} {\bibfnamefont {P.}~\bibnamefont
			{Chudzinski}}, \bibinfo {author} {\bibfnamefont {M.}~\bibnamefont {Gabay}}, \
			and\ \bibinfo {author} {\bibfnamefont {T.}~\bibnamefont {Giamarchi}},\ }\href
			{\doibase 10.1103/PhysRevB.78.075124} {\bibfield  {journal} {\bibinfo
			{journal} {Phys. Rev. B}\ }\textbf {\bibinfo {volume} {78}},\ \bibinfo
			{pages} {075124} (\bibinfo {year} {2008})}\BibitemShut {NoStop}%
	\bibitem [{\citenamefont {Nishimoto}\ \emph {et~al.}(2009)\citenamefont
			{Nishimoto}, \citenamefont {Jeckelmann},\ and\ \citenamefont
			{Scalapino}}]{ni.je.09}%
			\BibitemOpen
			\bibfield  {author} {\bibinfo {author} {\bibfnamefont {S.}~\bibnamefont
			{Nishimoto}}, \bibinfo {author} {\bibfnamefont {E.}~\bibnamefont
			{Jeckelmann}}, \ and\ \bibinfo {author} {\bibfnamefont {D.~J.}\ \bibnamefont
			{Scalapino}},\ }\href {\doibase 10.1103/PhysRevB.79.205115} {\bibfield
			{journal} {\bibinfo  {journal} {Phys. Rev. B}\ }\textbf {\bibinfo {volume}
			{79}},\ \bibinfo {pages} {205115} (\bibinfo {year} {2009})}\BibitemShut
			{NoStop}%
	\bibitem [{\citenamefont {Nishimoto}\ \emph {et~al.}(2010)\citenamefont
			{Nishimoto}, \citenamefont {Jeckelmann},\ and\ \citenamefont
			{Scalapino}}]{ni.je.10}%
			\BibitemOpen
			\bibfield  {author} {\bibinfo {author} {\bibfnamefont {S.}~\bibnamefont
			{Nishimoto}}, \bibinfo {author} {\bibfnamefont {E.}~\bibnamefont
			{Jeckelmann}}, \ and\ \bibinfo {author} {\bibfnamefont {D.}~\bibnamefont
			{Scalapino}},\ }\href {\doibase https://doi.org/10.1016/j.physc.2009.11.104}
			{\bibfield  {journal} {\bibinfo  {journal} {Physica C: Superconductivity and
			its Applications}\ }\textbf {\bibinfo {volume} {470}},\ \bibinfo {pages} {S53
			} (\bibinfo {year} {2010})},\ \bibinfo {note} {proceedings of the 9th
			International Conference on Materials and Mechanisms of
			Superconductivity}\BibitemShut {NoStop}%
	\bibitem [{\citenamefont {Bounoua}\ \emph {et~al.}(2019)\citenamefont
			{Bounoua}, \citenamefont {Mangin-Thro}, \citenamefont {Jeong}, \citenamefont
			{Saint-Martin}, \citenamefont {Pinsard-Gaudart}, \citenamefont {Sidis},\ and\
			\citenamefont {Bourges}}]{bo.ma.09u}%
			\BibitemOpen
			\bibfield  {author} {\bibinfo {author} {\bibfnamefont {D.}~\bibnamefont
			{Bounoua}}, \bibinfo {author} {\bibfnamefont {L.}~\bibnamefont
			{Mangin-Thro}}, \bibinfo {author} {\bibfnamefont {J.}~\bibnamefont {Jeong}},
			\bibinfo {author} {\bibfnamefont {R.}~\bibnamefont {Saint-Martin}}, \bibinfo
			{author} {\bibfnamefont {L.}~\bibnamefont {Pinsard-Gaudart}}, \bibinfo
			{author} {\bibfnamefont {Y.}~\bibnamefont {Sidis}}, \ and\ \bibinfo {author}
			{\bibfnamefont {P.}~\bibnamefont {Bourges}},\ }\href@noop {} {\enquote
			{\bibinfo {title} {Loop currents in two-leg ladders cuprates},}\ } (\bibinfo
			{year} {2019}),\ \Eprint {http://arxiv.org/abs/1912.07757} {arXiv:1912.07757
			[cond-mat.str-el]} \BibitemShut {NoStop}%
	\bibitem [{\citenamefont {Lichtenstein}\ and\ \citenamefont
			{Katsnelson}(2000)}]{li.ka.00}%
			\BibitemOpen
			\bibfield  {author} {\bibinfo {author} {\bibfnamefont {A.~I.}\ \bibnamefont
			{Lichtenstein}}\ and\ \bibinfo {author} {\bibfnamefont {M.~I.}\ \bibnamefont
			{Katsnelson}},\ }\href {\doibase 10.1103/PhysRevB.62.R9283} {\bibfield
			{journal} {\bibinfo  {journal} {Phys. Rev. B}\ }\textbf {\bibinfo {volume}
			{62}},\ \bibinfo {pages} {R9283} (\bibinfo {year} {2000})}\BibitemShut
			{NoStop}%
	\bibitem [{\citenamefont {Kotliar}\ \emph {et~al.}(2001)\citenamefont
			{Kotliar}, \citenamefont {Savrasov}, \citenamefont {P\'alsson},\ and\
			\citenamefont {Biroli}}]{ko.sa.01}%
			\BibitemOpen
			\bibfield  {author} {\bibinfo {author} {\bibfnamefont {G.}~\bibnamefont
			{Kotliar}}, \bibinfo {author} {\bibfnamefont {S.~Y.}\ \bibnamefont
			{Savrasov}}, \bibinfo {author} {\bibfnamefont {G.}~\bibnamefont {P\'alsson}},
			\ and\ \bibinfo {author} {\bibfnamefont {G.}~\bibnamefont {Biroli}},\ }\href
			{\doibase 10.1103/PhysRevLett.87.186401} {\bibfield  {journal} {\bibinfo
			{journal} {Phys. Rev. Lett.}\ }\textbf {\bibinfo {volume} {87}},\ \bibinfo
			{pages} {186401} (\bibinfo {year} {2001})}\BibitemShut {NoStop}%
	\bibitem [{\citenamefont {Liebsch}\ \emph {et~al.}(2008)\citenamefont
			{Liebsch}, \citenamefont {Ishida},\ and\ \citenamefont {Merino}}]{li.is.08}%
			\BibitemOpen
			\bibfield  {author} {\bibinfo {author} {\bibfnamefont {A.}~\bibnamefont
			{Liebsch}}, \bibinfo {author} {\bibfnamefont {H.}~\bibnamefont {Ishida}}, \
			and\ \bibinfo {author} {\bibfnamefont {J.}~\bibnamefont {Merino}},\ }\href
			{\doibase 10.1103/PhysRevB.78.165123} {\bibfield  {journal} {\bibinfo
			{journal} {Phys. Rev. B}\ }\textbf {\bibinfo {volume} {78}},\ \bibinfo
			{pages} {165123} (\bibinfo {year} {2008})}\BibitemShut {NoStop}%
	\bibitem [{\citenamefont {S\'en\'echal}(2015)}]{sene.15}%
			\BibitemOpen
			\bibfield  {author} {\bibinfo {author} {\bibfnamefont {D.}~\bibnamefont
			{S\'en\'echal}},\ }in\ \href@noop {} {\emph {\bibinfo {booktitle} {Many-Body
			Physics: From Kondo to Hubbard}}},\ Vol.~\bibinfo {volume} {5},\ \bibinfo
			{editor} {edited by\ \bibinfo {editor} {\bibfnamefont {E.}~\bibnamefont
			{Pavarini}}, \bibinfo {editor} {\bibfnamefont {E.}~\bibnamefont {Koch}}, \
			and\ \bibinfo {editor} {\bibfnamefont {P.}~\bibnamefont {Coleman}}}\
			(\bibinfo  {publisher} {Forschungszentrum J\"ulich},\ \bibinfo {year}
			{2015})\ pp.\ \bibinfo {pages} {13.1--13.22}\BibitemShut {NoStop}%
	\bibitem [{\citenamefont {Caffarel}\ and\ \citenamefont
			{Krauth}(1994)}]{ca.kr.94}%
			\BibitemOpen
			\bibfield  {author} {\bibinfo {author} {\bibfnamefont {M.}~\bibnamefont
			{Caffarel}}\ and\ \bibinfo {author} {\bibfnamefont {W.}~\bibnamefont
			{Krauth}},\ }\href {\doibase 10.1103/PhysRevLett.72.1545} {\bibfield
			{journal} {\bibinfo  {journal} {Phys. Rev. Lett.}\ }\textbf {\bibinfo
			{volume} {72}},\ \bibinfo {pages} {1545} (\bibinfo {year}
			{1994})}\BibitemShut {NoStop}%
	\end{thebibliography}

%
	
\end{document}